\documentclass[useAMS,usenatbib]{mn2e}
\usepackage{graphicx}
\usepackage{float}
\title{Statistical Mechanics of 1d Self-Gravitating Systems: The Core-Halo Distribution}
\author[T. N. Teles, Y. Levin and R. Pakter]{T. N. Teles$^{1}$\thanks{E-mail: teles@if.ufrgs.br}, 
Y. Levin$^{1}$\thanks{E-mail: levin@if.ufrgs.br} and R. Pakter$^{1}$\thanks{E-mail:pakter@ufrgs.br}\\
\\
$^{1}$ Instituto de F\'{\i}sica, Av. Bento Gon\c{c}alves, 9500, Caixa Postal 15051, 91501-970 - 
Porto Alegre, RS, Brazil}

\begin{document}

\date{ }

\pagerange{\pageref{firstpage}--\pageref{lastpage}} \pubyear{2011}

\maketitle

\label{firstpage}

\begin{abstract}
We study, using both theory and simulations, a system of self-gravitating sheets.
A new statistical mechanics theory --- free of any adjustable parameters --- is derived to quantitatively 
predict the final stationary state 
achieved by this system after the process of collisionless relaxation is complete. 
The theory shows a very good agreement with the numerical simulations.  The model sheds new light on
the general mechanism of relaxation of self-gravitating systems and may help us to 
understand cold matter distribution in the universe.
\end{abstract}


\section{Introduction}

One of the most fundamental questions of astrophysics concerns the mass distribution
after a self-gravitating system has relaxed to equilibrium.
This problem, posed more than 6 decades ago, still remains unresolved, see
the recent papers \citep{lili2, lili1} and references therein. In the astrophysical context, 
the answer to this question may help to shed light
on a lot of intriguing theoretical puzzles, such as the physical mechanism 
responsible for the regularities observed in the light profile of elliptical galaxies and
the mass distribution in the dark matter halos.

The classical simulations of \citet{nava1,nava2,nava3} have produced the density profiles of 
dark matter halos that no present theory is capable of explaining, 
see \citep{camm,hohl1,lb1,hohl2,cup1,shu,white,saslaw} 
and references therein. The scope of the problem extends from the foundations of statistical
mechanics to the large scale evolution of the universe, \citep{padrep1}.
The purpose of the present paper is to construct a statistical theory 
that is able to successfully predict both the mass and the 
velocity distributions of a self-gravitating system
after it has completed the process of collisionless relaxation. In this respect 
the One Dimensional Self-Gravitating Model (ODSGM) of interacting mass sheets, 
has proven very useful for understanding both the stellar dynamics and the
cosmological models, see (Wright, Miller \& Stein 1982; \citealt{mathur,joy3a,joy3b,mill1,joy1}) and references therein.
Although much simpler than the real three dimensional 
gravity, this model contains the essence of 
the gravitational problem: long-range potential~\citep{campa}
and  collective motion damped by particle-wave interactions which leads to collisionless 
relaxation, \citep{levin,levin2,teles,pakter} and references therein.
The ODSGM  is particularly convenient because the sheet-sheet interaction potential is not singular, 
so that the model can be easily simulated~\citep{noullez}.
At the same time it contains much of the same 
complicated statistical mechanics (Campa et al. 2009) 
of other systems with long-range forces: 
broken ergodicity and lack of mixing, 
which also plague gravity in three dimensions.

Long before the empirical works of Navarro et al. (1995, 1996, 1997)
Lynden-Bell (LB) proposed a statistical theory of self-gravitating systems 
based on the Vlasov equation~\citep{lb1}. The LB theory
has become known as the Theory of Violent Relaxation. One of the 
fundamental assumptions of LB was that violent relaxation leads to an efficient phase-space mixing. 
The subsequent simulations, however, have shown that this is not the case (\citealt{bindoni}; 
\citealt{yama}; Levin et al. 2008a,b; \citealt{teles,nava4}).  
Therefore, a new approach is needed if one wants to theoretically understand the structure of 
self-gravitating systems in equilibrium.
 
We will work 
in thermodynamic limit --- the number of sheets $N$ diverges, the mass of each sheet $m$ goes to zero, 
while the total mass $M$
of the system remains fixed, $m N=M$.
The advantage of working with 1d system is that the gravitational potential is 
unbounded from above and bounded from below.
As a consequence of this  --- after a sufficient large  time $\tau_\times$, known as the Chandrasekhar 
time --- a finite self-gravitating system must 
relax to the usual Maxwell-Boltzmann (MB) equilibrium~\citep{rib}.  
In the limit $N \rightarrow \infty$, the Chandrasekhar time
diverges and the system becomes trapped in an out-of-equilibrium stationary 
state~\citep{mill2,reidl,tsuchiya,tsuchiya2,tsuchiya3,yawn,teles,muka1}. 
Although 3d self-gravitating systems never relax to MB equilibrium they also become trapped in a
quasi-Stationary State qSS~(Levin et al. 2008a,b). 
In fact, for real 3d self-gravitating systems, such as elliptical galaxies,
the life time of the qSS is larger than their proper life span~\citep{tremaine}. For ODSGM it has been observed that qSS state has a very 
long life-span even for a small number of particles~\citep{mill2}.

This letter is organized as follows. In section~\ref{Model} we introduce the 1d gravitational sheet model. 
In section~\ref{Vlasov} we comment briefly on its general dynamics
and present the results of numerical simulation. In section
~\ref{Envelope} we derive the equation which is capable, over a short time span, to accurately 
describe the oscillations of the root-mean-square of particle positions.   
In section~\ref{Corehalo} we present the statistical-mechanics theory that is able to accurately 
predict both the velocity and the mass distribution in equilibrium, without any adjustable parameters. 
The more involved derivation of the envelope equation is left to the appendix of the paper, ~\ref{Enve}.

\section[]{The Model's Description}
\label{Model}

The ODSGM consists of $N$ sheets of mass $m_i$ in the $y-z$ plane, free to move along the $x$-axis.
To simplify the calculations, we suppose that 
all the sheets have the same mass $m_i=m=M/N$, where $M$ is the total mass of the system. 
The sheets interact through the one dimension gravitational potential that satisfies the 
Poisson equation
\begin{equation}
\label{eqpoisson}
  \nabla^2\psi(x,t)=4\pi G \rho(x,t) \,,
\end{equation}
where $G$ is the gravitational constant and $\rho(x,t)$ is the mass density. 

We define dimensionless 
variables by scaling mass, lengths, velocities, the potential, the mass density and the 
energy with respect to $M$, $L_0$ (an arbitrary length scale), $V_0 =\sqrt {2\pi G M L_0}$, 
$\psi_0 = 2\pi G M L_0$, $\rho_0=M/2 L_0$ and $E_0 = M V_0^2 = 2\pi G M^2 L_0$, 
respectively. All these scales correspond to setting $G=M=1$ and to defining 
the dynamical time scale as,
\begin{equation}
\label{timescale}
\tau_D=(4\pi G \rho_0)^{-1/2}.
\end{equation}
It is then easy to show that a particle (sheet) at the origin with a mass density 
$\rho(x)=\delta(x)$ creates a long-range gravitational potential,
\begin{equation}
\psi(x)=|x|\;.
\end{equation}
Systems with long-range interaction 
are very peculiar because in the  thermodynamic limit the collision duration time diverges, and the dynamical 
evolution of the system is governed exactly by the collisionless Boltzmann (Vlasov) equation, \citep{braun}
\begin{equation}
\label{eqvlasov}
  \frac{D f}{D {\rm t}}=\frac{\partial f}{\partial {\rm t}}+{\rm v}\frac{\partial f}{\partial {{\rm x}}}-
\nabla \psi({\rm x,t})\frac{\partial f}{\partial {\rm v}}=0,
\end{equation}
where $f({\rm x,v,t})$ is the one particle distribution function, so that $\rho({\rm x, t})=\int f({\rm x,v,t}){\rm dv}$.
Clearly real self-gravitating systems do not rigorously obey the $N\rightarrow\infty$ limit, nevertheless, in the
astrophysical context the 
number of "particles" is usually very large, so that it constitutes a very good approximation.

The Vlasov equation has an infinite number of invariants called the Cassimirs \citep{chavani}. 
Two of these are mass and energy,
\begin{equation}
\label{eqvolconserv}
\int {\rm dx}\,{\rm dv}\;f({\rm x,v,t}) = 1\;,
\end{equation}
\begin{equation}
\label{eqeneconserv}
\int {\rm dx}\,{\rm dv}\; \left(\frac{{\rm v}^2}{2}+\frac{\psi({\rm x})}{2}\right)\; f({\rm x,v,t}) = {\cal E}_0 \,,
\end{equation}
respectively, where ${\cal E}_0$ is the initial energy. Linear ${\it momentum}$ is also a 
conserved quantity. However, by the symmetry of the problem and without loss of generality, it can be set to zero.
In fact, any local functional of the distribution function is a Cassimir invariant of the Vlasov dynamics.

\section{Vlasov Dynamics}
\label{Vlasov}
Unlike collisional systems which after a very short time relax to the Maxwell-Boltzmann distribution,
the time evolution of the Vlasov equation never ends.  Instead it progresses to smaller and smaller length scales.
In practice, however, both experiments and simulations have only a finite resolution. 
It is in this, coarse-grained, sense that we say that the Vlasov dynamics 
evolves to the stationary state.  
Unlike the usual MB equilibrium, however, the stationary state reached through the process of
collisionless relaxation depends explicitly on the initial particle distribution.

\subsection{Numerical Simulations}
\label{simulations}

Numerical solution of the Vlasov equation is very complicated.  One alternative is to perform directly the  
N-body simulation in which all particles evolve over time according to the 
usual Newton equations of motion~\citep{noullez}.  
We consider a ODSGM in which sheets are initially distributed in accordance with the water-bag distribution,
\begin{equation}
\label{eqwaterbag}
f_0(x,v) = \eta \Theta(x_m-|x|)\Theta(v_m-|v|)
\end{equation}
where $\Theta$ is the Heaviside step function and $\eta=(4\ x_m v_m)^{-1}$ is the normalization 
constant obtained from Eq.~(\ref{eqvolconserv}). 
The values $x_m$ and $v_m$ are the limits of the distribution function in the phase space, so that
without loss of generality we set $x_m=1$. Due to the symmetry of the distribution, the total
linear {\it momentum} is null and the only quantity other than the total mass to be conserved is
the total energy ${\cal E}_0=v_m^2/6 + 1/3$.

The halo formation is related to the initial condition through the only dimensionless parameter that
remains after the rescaling of section \ref{Model} --- the virial number, ${\cal R}=2T/U$,
where $T$ is the total kinetic and $U$ is the potential energy.
When the virial condition $2T=U$ is not satisfied, ${\cal R} \ne 1$, the imbalance between 
the kinetic and the potential energies causes the system to develop collective oscillations which are 
then damped by particle-wave interactions, leading to the core-halo phase separation, Fig.1. 
\begin{figure}
\begin{center}
\includegraphics[width=8.cm]{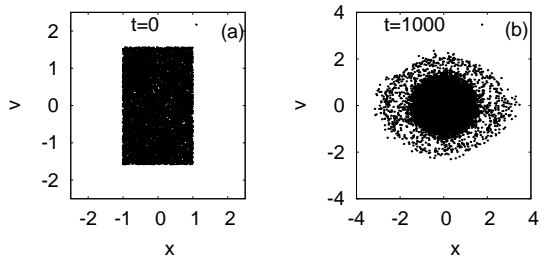}
\caption{The  initial (a) and final (b) phase space of $10^4$ self-gravitating sheets initially distributed 
according to the water-bag distribution with the virial number ${\cal R}_0=2.5$.}
\end{center}
\label{figlbch}
\end{figure}
Our interest in studying the distributions with ${\cal R}\neq 1$ arises from the fact that 
there is not even a qualitative theory that is able to account 
for the mass and velocity distributions for such systems \citep{yamash}. 

\section{The test particle model}
\label{Envelope}

Systems with long-range forces display collective motion 
such as Langmuir waves in plasmas  or Jeans waves in gravitational systems, \citet{tremaine}.
Since there are no collisions, 
interaction of individual particles with the density waves is the mechanism that responsible for the transfer
of energy. The particle-wave interaction dampens the density waves while at
the same time transfers large amounts of energy to the individual particles.  The interaction is similar
to a surfer "catching" a wave  ---  some particles can gain a large amount of energy from the bulk oscillations 
to escape from the gravitational cluster and move to high  energy regions of the phase space.  
These particles will then form a
tenuous halo that will surround the dense low-energy core.

To study the oscillation of the initial gravitational cluster, 
we define the {\it envelope} as $x_e(t)\equiv \sqrt{3<x(t)^2>}$.  Note that with this definition, 
at $t=0$ the envelope is exactly the same as the extent of the initial particle distribution
$x_e(0)=x_m$.  For sufficiently short times, most particles will still remain within the limits set by $x_e(t)$.
Differentiating $x_e(t)$ twice with respect to time, we find an approximate dynamical equation satisfied by  $x_e(t)$,
\begin{equation}
\label{eqenvel}
{\ddot x}_e(t)=\frac{{\cal R}_0}{x_e(t)}-1
\end{equation}
where ${\cal R}_0$ is the virial number of the initial distribution (the derivation is presented in 
Appendix~\ref{Enve}). We shall call this the "envelope equation" \citep{wang98, dav01}. 
Equation~\ref{eqenvel} enables us to study the dynamics of a test particle $x_i$, subject to the
oscillating potential produced by a uniform density distribution delimited by $-x_e(t) \le x \le x_e(t)$. 
A test particle moving in this potential will then evolve according to 
\begin{eqnarray}
\label{eqpoinc}
\ddot{x}_i(t)&=&
\left\{
\begin{array}{l}
-\frac{x_i(t)}{x_e(t)}\>\, $for$ \ \,|x_i(t)|\le x_e(t)
\\
\\
-$sgn$\left[x_i(t)-x_e(t)\right] \>\, \ $for$ \ \,|x_i(t)|\ge x_e(t)
\end{array}
\right.
\end{eqnarray}
where $x_e(t)$ is given by Eq.~(\ref{eqenvel}) and $sgn$ is the sign function. 
\begin{figure}
\begin{center}
\includegraphics[width=8.cm]{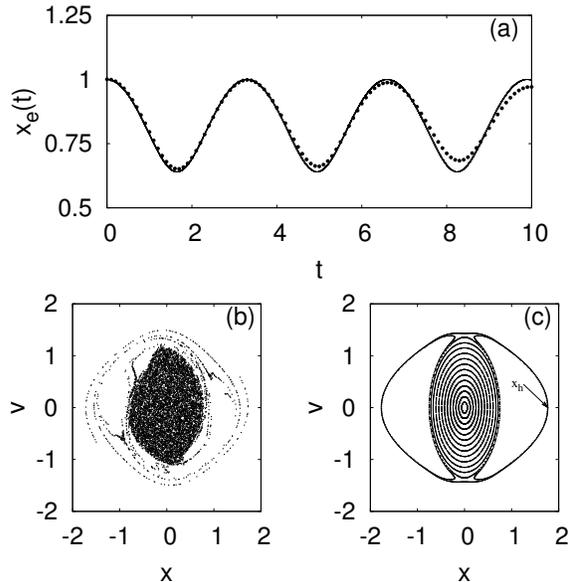}
\caption{Test particle dynamics as compared to the $N$-body dynamics 
simulation: (a) corresponds to the evolution of
$x_e(t)$ obtained from the $N$-body simulation (dots) while the solid line is 
calculated using the envelope equation. The time is in units of $\tau_D$.
(b) Represents the
phase space after the first few oscillations and (c) is the Poincar\'e plot of the phase space produced by the
test particles. Note the appearance of the resonance islands both in the N-body and test particle simulation.
The test particle model allows us to accurately calculate the location of the resonant orbit with the 
maximum extent $x_h$. 
This will determine the
maximum energy attained by the halo particles. At $t=0$, $15$ test particles were uniformly distributed from the
center to slightly beyond the maximum core radius $x_m$ with initial velocity $0$.}
\end{center}
\label{figenve}
\end{figure}
In Fig. 2 we compare the complete $N$-body simulation with the test
particle trajectories obtained from the numerical
solution of Eqs.~(\ref{eqenvel}) and (\ref{eqpoinc}).
The good agreement between the two reveals that for short times the envelope equation~\ref{eqenvel}
is very accurate for describing the initial oscillations of the ODSGM.  The formation of
halo also occurs on a very short time scale  --- few oscillations of the envelope.  Indeed, 
from Figs. 2 (b), (c), we see that
the test particle model allows us to accurately locate the resonant orbits~\citep{gluckstern}, which also
delimit the extent of the halo in the full $N$-body simulations. Thus, the test particle model allows us to calculate
the maximum energy --- corresponding to the resonant orbit --- attained by any of the halo particles.  This energy 
will be used in the next Section to construct the core-halo distribution function.

\section{The Core-Halo Statistical Theory}
\label{Corehalo}

The Jeans theorem states that ``any steady-state solution 
of the collisionless Boltzmann equation depends on the phase-space coordinates 
only through integrals of motion of the given potential, and any function of 
the integrals yields a steady-state solution of the collisionless Boltzmann equation'',~\citep{tremaine}. 
Thus, any function of energy is a solution to the Vlasov equation. In particular,
the Maxwell-Boltzmann  distribution (MB) is also a stationary solution of the Vlasov equation.
However, it is important to remember that MB is not a global attractor of the Vlasov dynamics --- 
an arbitrary initial distribution of particles will not converge to the MB distribution, as happens for
systems governed by short-range forces.
In this paper we are interested in calculating the distribution to which the system will evolve,
starting from an arbitrary water-bag-like initial condition, after the process of collisionless relaxation is
completed.

We first observe that the Vlasov equation can be interpreted as a convective derivative of density over 
the phase-space, so that the distribution function evolves as an incompressible fluid. 
In the case of initial water-bag distribution this means that the phase-space density cannot exceed that of $f_0$, 
$f({\rm x,v,t})\le \eta$. 
The mechanism of core-halo phase separation now becomes clear. Some particles enter 
in resonance with the rms bulk oscillations. They gain a lot of energy and escape from the main cluster, forming
a tenuous halo.  At the same time, evaporation of highly energetic particles dampens the core oscillations.  This is
similar to the process of evaporative cooling.  The oscillations will stop when all of the free energy is exhausted.
However, because of the constraint on the maximum density imposed by the Vlasov dynamics, the system
cannot freeze --- collapse to the minimum of the potential energy.  Instead, the particle distribution of the
core approaches that of a fully degenerate Fermi gas.  With this physical insight in mind, we now 
propose an {\it ansatz} solution to the Vlasov equation: 
\begin{eqnarray}
\label{corehaloeq}
f_{ch}({\rm x, v})&=&\eta \Theta\left[\epsilon_F-\epsilon\right]
+\chi \Theta\left[\epsilon_h-\epsilon\right]\Theta\left[\epsilon-\epsilon_F\right]
\end{eqnarray}
where  $\epsilon_F$ is the Fermi energy,  $\epsilon_h$  is the halo (resonance) energy, and
$\epsilon(x,v)=v^2/2+\psi(x)$ is the one particle energy. The halo energy is obtained
simply by taking the most extended position of test particle $x_h$ when it crosses 
the  $v=0$ axis,  in others words $\epsilon_h=|x_h|$. 
The Fermi energy, $\epsilon_F$, delimits the extent of the core region, and $\chi$ measures
the phase space density inside the halo.  Both of these parameters are calculated self-consistently
from the norm and energy conservation equations,  Eqs.~(\ref{eqvolconserv}) and (\ref{eqeneconserv}).

Substituting Eq.~\ref{corehaloeq} into the Poisson equation and
integrating the core-halo distribution function over velocity, the dimensionless Poisson
equation becomes
\begin{eqnarray}
\label{eq poisson2}
\frac{d^2\psi}{d{\rm x}^2}=2\sqrt{2}\left\{
\begin{array}{l}
(\eta-\chi) \sqrt{\epsilon_F - \psi}+\chi \sqrt{\epsilon_h-\psi}\>\, \> \> $for$\> \> \psi \le  \epsilon_F  \\
\\
\chi \sqrt{\epsilon_h-\psi}\>\, \> \>$for$\> \> \epsilon_F \le \psi \le \epsilon_h \\
\\
0\>\,\> \> $for$\> \> \psi \ge \epsilon_h \,.
\end{array}
\right.
\end{eqnarray}

Solving this equation numerically and using the constraints Eqs.~(\ref{eqvolconserv}) and (\ref{eqeneconserv})
yield the complete solution for the one-particle distribution function. 
In Fig. 3 we compare the theoretically calculated density
and velocity distribution functions to the full $N$-body simulation. 
To obtain the distributions in the simulations, the system is evolved
for $1000$ dynamical times until the stationary state is achieved.
To get good statistics the histograms are constructed by binning the
particle positions and velocities at four different times.
An excellent agreement is found between the theory and the
simulations, {\it without any adjustable parameters}.  Note that the agreement
is very good for both the virial numbers above and below ${\cal R}_0=1$.  In the same figure we also plot,
with the dashed lines, the initial density and the velocity distributions.  
It is clear 
that during the process of 
collisionless relaxation the system moves far from the initial condition.  It is important to keep in mind that
although the system has relaxed to the stationary state, this state does not correspond to the thermodynamic 
equilibrium.
In particular, the velocity distribution does not have the equilibrium Maxwell-Boltzmann form.  This also shows that
one needs to take a special care when modeling collisionless system using hydrodynamic-like equations, since there 
might not be thermodynamic equilibrium even locally. 

\begin{figure}
\begin{center}
\includegraphics[width=8.cm]{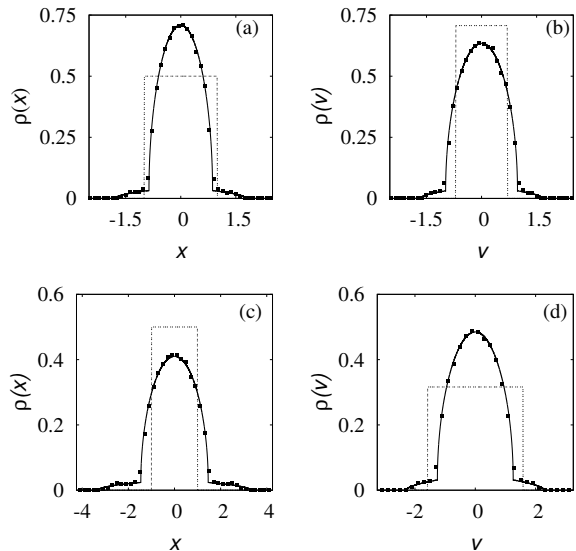}
\caption{Final position (a) and velocity (b) distributions for initial
water-bag  with ${\cal R}_0=0.5$ at t=1000 $\tau_D$. (c) 
and (d) are the stationary distribution for the initial water-bag with ${\cal R}_0=2.5$. The dotted lines 
are the initial distributions.}
\end{center}
\label{figtheory}
\end{figure}

\section{Conclusions}
\label{conclusion}

We have presented a theory which allows us to {\it a priori} predict the density and the velocity distribution
functions for a system of gravitationally interacting sheets.  The theory is quantitatively accurate without
any fitting parameters. The theory and the simulations show that in the thermodynamic limit this system does 
not evolve to the usual MB equilibrium.  Instead it becomes trapped in a non-ergodic, non-mixing state ---  
once formed, the halo never again equilibrates with the core in the $N \rightarrow \infty$ limit.  

Although in this paper we have studied only
water-bag-like initial distributions, the theory can be easily generalized to more complex 
functional forms \citep{teles2009}. The next step
in the development of the theory will  be to include a specific cosmological model to account for the expansion
of the universe.  If this can be achieved, the theory might shed new light on the mass distribution 
in the dark-matter halos.

\section*{Acknowledgments}
T.N.T. is grateful to Fernanda Benetti for help with writing the manuscript. The work was partially supported by the 
CNPq, Fapergs, INCT-FCx, 
and by the US-AFOSR under the grant FA9550-09-1-0283.

\appendix



\section{Envelope Equation}
\label{Enve}

We define the "envelope" as $x_e\equiv \sqrt{3<x^2>}$.  Differentiating twice with respect to time we find,
\begin{equation}
\label{dd}
\ddot {x}_e=\frac{3<x\ddot{x}>}{x_e}+\frac{3<\dot{x}^2>}{x_e}-\frac{9<x\dot{x}>^2}{x_e^3}.
\end{equation}
To simplify the first term, we suppose that the 
mass density oscillations are affine --- preserve the uniform mass distribution.   In this case,
\begin{equation}
<x\ddot{x}>=-<x\frac{d \psi}{{\rm dx}}>=-\frac{1}{2x_e(t)}\int_{-x_{e}(t)}^{x_e(t)}\frac{x^2}{x_e}{\rm dx}\,.
\end{equation}
Integrating, we find 
\begin{equation}
<x \ddot{x}>=-\frac{x_e(t)}{3}.
\end{equation}

The second term of eq. (\ref{dd}) involves only the mean values which, for short times, should remain close to those
obtained using the initial particle distribution function $f_0$,
\begin{eqnarray}
<\dot{x}^2>=\frac{1}{2v_m}\int_{-v_m}^{v_m}\dot{x}^2{\rm d\dot{x}}=\frac{v_m^2}{3}
\\ \nonumber
\\ \nonumber
<x\dot{x}>=\frac{1}{4\ x_m v_m}\int_{-x_m}^{x_m}x{\rm dx}\int_{-v_m}^{v_m}\dot{x}{\rm d\dot{x}}=0
\end{eqnarray}
Finally the envelope equation reduces to,
\begin{equation}
\label{envequation}
{\ddot x}_e(t)=\frac{{\cal R}_0}{x_e(t)}-1\;,
\end{equation}
where ${\cal R}_0=\frac{2 T_0}{U_0}=v_m^2$. Since $x_e(0)=x_m=1$ if ${\cal R}_0=1$ then $\ddot{x}_e(t)=0$, 
and the system does not develop macroscopic collective oscillations.  


\label{lastpage}

\end{document}